\documentclass[twocolumn,showpacs,preprintnumbers,amsmath,amssymb,aps,prl,footinbib,superscriptaddress]{revtex4}
\usepackage{graphicx}

\begin{document}

\title{Holographic Mean-Field Theory for Baryon Many-Body Systems}

\author{Masayasu Harada}
\affiliation{
Department of Physics, Nagoya University, Nagoya 464-8602, Japan
}
\author{Shin Nakamura}
\affiliation{
Department of Physics, Kyoto University, Kyoto 606-8502, Japan
}
\author{Shinpei Takemoto}
\affiliation{
Department of Physics, Nagoya University, Nagoya 464-8602, Japan
}

%\date{\today}

\begin{abstract}
We propose a mean-field approach to analyze many-body systems of fermions in the gauge/gravity duality. We introduce a non-vanishing
classical fermionic field in the gravity dual, which we call the holographic mean field for fermions. The holographic mean field takes account of the many-body dynamics of the fermions in the bulk. The regularity condition of the holographic mean field fixes the relationship between the chemical potential and the density
unambiguously. Our approach provides a new framework of gauge/gravity duality for finite-density systems of baryons in the confinement phase. 
\end{abstract}


\maketitle

Many-body physics of strongly-correlated fermions is one of the central subjects in modern physics. Strongly-correlated fermions appear in various places, such as the systems of cold atoms, high-$T_{c}$ superconductors, neutron stars and the quark-gluon plasma. However, the theoretical description of many-body systems of strongly-correlated fermions is still a challenge owing to their non-perturbative natures. One of the successful computational frameworks for strongly-interacting fermions is the lattice gauge theory, but the Monte Carlo simulation at finite baryon density suffers from the sign problem.

Recently, a great deal of attention has been paid to the AdS/CFT correspondence \cite{Maldacena:1997re,Gubser:1998bc,Witten:1998qj} 
because of its ability for non-perturbative computations and the absence of the sign problem. However, the current framework of the AdS/CFT correspondence
for finite baryon density still has room for improvement in the confinement phase. 
It can be summarized as follows: 1) 
dynamical baryons at finite density have not been satisfactorily incorporated, and 
2)
the Gubser-Klebanov-Polyakov-Witten (GKP-Witten) prescription~\cite{Gubser:1998bc,Witten:1998qj} to relate the baryon chemical potential to the baryon density has an ambiguity~\cite{Nakamura:2007nk}.

In this paper, we shall 
propose a new approach, a holographic mean-field theory for fermions, 
to resolve these problems. The self-consistent incorporation of dynamical baryons in our approach provides an eigenvalue equation which fixes the relationship between the baryon chemical potential and the baryon density unambiguously.

Let us revisit the above mentioned problems in detail below.
The quarks are realized as fundamental strings attached to the flavor brane in the gravity dual~\cite{Erdmenger:2007cm}.
Each end point of the strings carries a unit charge with respect to a $U(1)$ gauge field on the flavor brane, and the quark density can be evaluated from the ``electric field" in the radial direction at the boundary through the Gauss-law constraint.
The quark chemical potential is thus the time component of the $U(1)$ vector potential, $A_{0}$, at the boundary~\cite{Kim:2006gp}, which is conjugate to the electric field. 
This fits the conventional GKP-Witten prescription: the boundary value of $A_{0}$ gives the source (the chemical potential); the expectation value of the conjugate operator (the density) is given by the derivative of $A_{0}$ at the boundary with respect to the radial direction. These quantities are boundary conditions for the Gauss-law constraint which may be chosen arbitrarily.
However, another physical condition {\em in the bulk}, i.e., the bulk condition, fixes the ambiguity in the relationship between them whenever the GKP-Witten prescription works. To the best of the authors' knowledge, only the example where the bulk condition is yet to be known is the finite density systems of baryons
in the confinement phase.

The origin of the ambiguity can be understood in the following way.
The global $U(1)_{B}$ ($U(1)$-baryon) symmetry is enhanced to the $U(1)$ gauge symmetry in the gravity dual. 
The constant shift of the boundary value of $A_{0}$ can be a gauge transformation which does not alter the physics in the gravity dual whereas the constant shift of the chemical potential has definite physical meaning in the boundary theory. This contradiction is resolved when the flavor brane intersects the horizon: $A_{0}$ has to vanish at the horizon to make the one-form well-defined there~\cite{Braden:1990hw,Kobayashi:2006sb}. 
This removes the ambiguity of the constant shift of the boundary value of $A_{0}$.
However, we cannot employ this prescription if the flavor brane does not intersect the horizon:
we need another bulk condition from which the ambiguity in the constant shift of $A_{0}$ is eliminated. 

More serious problem is the difficulty to incorporate many-body systems of dynamical baryons. In the phase where the flavor brane is away from the horizon, quarks are confined. 
The $U(1)_{B}$ charge is carried by baryons in this case. 
We assume that the number of the colors, $N_{c}$, is odd: the baryons are fermions. 
A single baryon is holographically realized as a baryon vertex \cite{Gross:1998gk} which consists of $N_{c}$ fundamental strings and a D-brane wrapped on a compact subspace. This is a fermionic object that lives inside the bulk of the dual geometry. 

In order to prepare a system of finite density of baryons, we need to deal with the many-body physics of the baryon vertices. We expect that their distribution along the radial direction in the dual geometry is non-trivially determined by their dynamics. However, it is very difficult to solve the many-body dynamics of the baryon vertices within the current technology of superstring theory: we need a good approximation.
Although they can be described as solitons (or Skyrmion-like objects)~\cite{Nawa:2006gv,HRYY,Hata:2007mb} on the flavor brane, the many-body physics of solitons on the curved spacetime is still hard to solve \cite{Rozali:2007rx}: we need further simplification.
In most of the literature \cite{Bergman:2007wp}, many-body systems of baryons in the gravity dual have been constructed just by superposing the baryon vertices at the same location in the radial direction.
However, the dynamics of the baryon vertices is not fully taken into account in this method: the system is over-simplified.

One way to overcome the difficulty is to approximate the baryon vertex as a point-particle fermion rather than the wrapped D-brane with $N_{c}$ strings,
and to describe their many-body dynamics in terms of  
a Dirac field~\cite{HIY,HRYY,Kim:2007xi} on the dual geometry. Indeed, this approximation makes sense at the strong coupling where the size of the soliton becomes small \cite{HRYY,Hata:2007mb}.
In this paper, we proceed along this approximation. We construct a system of finite
baryon density by using the non-vanishing bulk Dirac field, which we
call the {\em holographic mean field for baryons}.

Since the baryon vertex carries the $U(1)$ charge,
the holographic mean field is also requested to carry it.
Then the mean field induces non-zero expectation values of the fermion {\em bi-linear} operators, such as the charge density which is detected by the derivative of the $U(1)$ gauge field at the boundary. 
The dynamical distribution of the baryon vertices along the radial direction is represented by the configuration of the mean field.
Furthermore, the coupled equations of motion for the mean field and the $U(1)$ gauge field give an eigenvalue equation from which the chemical potential is unambiguously fixed.
It will also be shown that the chemical potential at the zero-density limit automatically agrees with the mass of the baryon. Thus, the problems 1) and 2) are resolved in harmony in our approach. 

Let us consider a five-dimensional (5D) dual geometry without an event horizon whose metric depends only on the radial direction, for example. It is convenient, although not necessary, to make a coordinate transformation after which the geometry is given by the conformally flat metric,
\begin{equation}
ds_{5d}^2 = H(w) ( \eta_{\mu\nu} dx^{\mu} dx^{\nu}-dw^2  ) \, , 
\label{5dmetric}
\end{equation}
where $\eta_{\mu\nu}=\mbox{diag}(+,-,-,-)$ and $w$ is the radial direction.
An example of $H(w)$ is given in~\cite{HRYY}. 
We introduce a 5D Dirac spinor $\Psi$ which interacts with a bulk $U(1)$ gauge field $A_M$ on this geometry. 
The actions for them are
\begin{align}
S_\Psi &= \int d^4x dw \left[ i \bar \Psi \Gamma^M (\partial_M - i q A_M) \Psi - m_5(w)  \bar \Psi \Psi \right] \,,  \notag \\
S_A &= \int d^4 x dw \mathcal{L}_A \,, \label{nuaction}
\end{align}
where $q$ and $m_5(w)$ denote the charge and the 5D mass of the fermion, respectively. $\Gamma^M$ are written in terms of the four dimensional (4D) Dirac matrices as $\Gamma^\nu = \gamma^\nu$ ($\nu = 0,1,2,3$) and $i \Gamma^w = \gamma^5$.
The possible boundary terms and the counter terms are omitted here.
Note that the Dirac field is on the curved spacetime (\ref{5dmetric}), but
the vielbein and the spin connection are absorbed by a suitable field redefinition of the fermion field~\cite{HRYY}. 
The $w$ dependence of $m_5(w)$ reflects the curvature of the geometry. Typically, $m_5(w)$ diverges at the boundary, and its explicit example is given in \cite{HRYY}. 
We assume that $S_A$ is a functional of $\partial_N A_M$,
which may be given by a Dirac-Born-Infeld (DBI) action on the flavor brane, for example. 
However, its explicit form is not important at this stage.
We employ the probe approximation such that the fermion field and the gauge field do not modify the geometry: we regard $S= S_{\Psi} + S_A$ as the total action up to the constant action of the gravity sector.

We now explain how we formulate the holographic mean-field approach.
We work in the $A_w (x,w) = 0$ gauge, and replace the gauge field with the mean field as
\begin{equation}
A_0 (x,w) \to A_0 (w) \,, \quad A_i (x,w) \to A_i (w) \,,
\end{equation}
where $i=(1,2,3)$.
We impose
\begin{equation}
A_0 (w) \bigr|_\text{boundary} = \mu \,, \quad A_i (w) \bigr|_\text{boundary} = 0 \,,  \label{bcuv}
\end{equation}
where $\mu$ is the chemical potential associated with the fermion
charge. 
We also introduce the mean field for the fermion as 
\begin{equation}
\Psi (x,w) \to \Psi (w) \,.
\end{equation}
The non-vanishing mean field realizes
the dynamical distribution of the charge in $w$ direction, which is given by
\begin{equation}
\rho (w) = q \Psi^\dagger (w) \Psi (w) \,.
\end{equation}
The integration of $\rho (w)$ over $w$ gives the 4D charge density $n$.
In the present analysis, we switch off the fermionic source:
\begin{equation}
\Psi (w) \bigr|_\text{boundary} = 0 \,. \label{bcuvp}
\end{equation}
We also require that the mean fields are regular. 

Now $A'_0  \equiv \partial_w A_0$ is
given by the Gauss-law constraint,
\begin{equation}
\partial_w [\partial \mathcal{L}_\text{A}/\partial A'_0(w)] - q \Psi^\dagger (w) \Psi (w) = 0 \,, \label{aeq}
\end{equation}
and $A'_i  \equiv \partial_w A_i$ is given by
\begin{equation}
\partial_w [\partial \mathcal{L}_\text{A}/\partial A'_i(w)] - q \Psi^\dagger (w) \Gamma^0 \Gamma^i \Psi (w) = 0 \,. \label{aieq}
\end{equation}
The scalar potential $A_0 (w)$ is given by using $A'_0 (w)$:
\begin{equation}
A_0 (w) = \mu + \int^w_\text{boundary} d w' A'_0 (w') \equiv \mu + \tilde{A}_0 (w) \,.
\end{equation}
However, the boundary value $\mu$, that is the constant of integration,
is arbitrary at this stage.
In other words, whatever value of $\mu$ can give the same charge distribution.
This has been a problem in holographic models at finite density.
This problem will be resolved by solving the equation of motion for $\Psi (w)$:
\begin{equation}
\left[ i \Gamma^w \partial_w + q \Gamma^0 A_0 (w) + q \Gamma^i A_i (w) - m_5(w)  \right] \Psi(w) = 0 \, \label{Dirac}
\end{equation}
that depends on the boundary value of $A_0(w)$ explicitly.

Let us see how it works.
We write $\Psi (w)$ in terms of the two-component spinors $\Psi_\pm$ as
$\Psi (w) = (\Psi_+,\Psi_-)^T$
with the Dirac matrices given by
\begin{equation}
i \Gamma^w = \left( \begin{array}{cc}
  0 & 1 \\ 1 & 0 \\
 \end{array}\right)
\,,\ 
\Gamma^0 = \left( \begin{array}{cc}
  1 & 0 \\ 0 & -1 \\
\end{array}\right)
\,,\ 
\Gamma^i = \left( \begin{array}{cc}
  0 & \sigma^i \\ - \sigma^i & 0 \\
\end{array}\right)
\,.
\end{equation}
We can take $A_1(w) = A_2(w) = 0$, and hence $\sigma^3 \Psi_\pm = \Psi_\pm$ without loss of generality by virtue of the rotational invariance.
From (\ref{Dirac}), we obtain
\begin{align}
[\partial_w - q A_3(w)] \Psi_+ - [ m_5(w) + q \tilde{A}_0 (w) ] \Psi_- &= q \mu  \Psi_- \,, \notag  \\
[\partial_w + q A_3(w)] \Psi_- - [ m_5(w)  - q \tilde{A}_0 (w) ] \Psi_+ &= - q \mu  \Psi_+ \,. \label{peq1}
\end{align}

Solving (\ref{peq1}) coupled with (\ref{aeq}) and (\ref{aieq}), $q \mu$ is obtained as the eigenvalue for a given value of $n = \int dw \, q \Psi^\dagger (w) \Psi (w)$. The point here is that the eigenvalue equations (\ref{peq1}) are nonlinear equations after rewriting the gauge fields in terms of $\Psi_+$ and $\Psi_-$, and the amplitudes of $\Psi_+$ and $\Psi_-$ are not arbitrary if we specify $q \mu$. If we start with a given $q \mu$, the density $n$ is determined as a function of $q \mu$. 
%Note that $n$ determines the normalization of $\Psi (w)$ as well.
In this way, the holographic mean-field theory provides 
the equation of state (EOS): the relation between the 4D charge density $n$ and the chemical potential $\mu$ is determined dynamically. 
The new bulk condition to fix the ambiguity of $\mu$ in the GKP-Witten prescription 
is the regularity of the fermionic mean field under (\ref{bcuvp}).

Once the mean fields are obtained, the energy spectrum of the physical excitations can be computed by introducing the fluctuations $\psi (x,w), a_\nu (x,w)$ on top of the mean fields as
$
\Psi (x,w)=\Psi (w) + \psi (x,w)
$ and $
A_\nu(x,w)=A_\nu(w) + a_\nu(x,w)
$,
and solving the coupled equations of motion for the fluctuations.
One of the equations of motion is given by
\begin{align}
&[  i \Gamma^w \partial_w + \Gamma^0 (p_0 + q A_0 (w)) + \Gamma^3 q A_3(w) + \vec{\Gamma} \cdot \vec{p} \notag \\
&- m_5(w) ] \psi (p_0, \vec{p} ,w) + \Gamma^\nu \Psi (w) q a_\nu (p_0, \vec{p} ,w) = 0 \,.
\label{Diracf}
\end{align}

Let us discuss the zero density limit, $n \to +0$. 
(\ref{aeq}) and (\ref{aieq}) tell us that $\tilde{A}_0 (w) \to 0, A_3 (w) \to 0$, and (\ref{peq1}) becomes
\begin{align}
\partial_w \Psi_+ - m_5(w) \Psi_- &= q \mu \Psi_- \,, \notag \\
\partial_w \Psi_- - m_5(w) \Psi_+ &= - q \mu \Psi_+ \,. \label{peq3}
\end{align}
On the other hand, at $\mu = 0$, all the mean fields vanish, then (\ref{Diracf}) is reduced to be
\begin{align}
\partial_w \psi_+ - m_5(w) \psi_- &=  m_f \psi_- \,, \notag \\
\partial_w \psi_- - m_5(w) \psi_+ &= - m_f \psi_+ \,, \label{peq5}
\end{align}
where $\psi = (\psi_+, \psi_-)^T$ and we used $\Gamma^\nu p_\nu \psi  = m_f \psi$.

The eigenvalue $m_f$ of (\ref{peq5}) for the non-vanishing eigenfunction under the boundary condition $\psi \bigr|_\text{boundary} = 0$ is the mass of the fermion.
Therefore, the threshold of the chemical potential for the non-vanishing density given by (\ref{peq3}) coincides with the fermion mass, i.e., $q \mu \to m_f$ at the limit $n \to +0$.
We do not find any nontrivial solutions for $\Psi (w)$ for $\lvert q \mu \rvert < m_f$, where there is no formation of the Fermi surface.

Now we demonstrate how to obtain the EOS for baryon many-body systems.
We apply our method to the model given in \cite{HRYY} where the baryon field is introduced into the Sakai-Sugimoto model~\cite{SS}.
The boundaries in this model are located at $w = \pm w_\text{max}$, thus the boundary conditions for the mean fields are $A_0 (\pm w_\text{max}) = \mu$, $A_3 (\pm w_\text{max}) = 0$, $\Psi (\pm w_\text{max}) = 0$.
Here $\mu$ is the baryon chemical potential and $q = 1$.
We assume that the theory is invariant under the 5D parity $(t, \mathbf{x}, w) \to (t, -\mathbf{x}, -w)$~\cite{SS}. 
%\footnote{See, e.g., \cite{SS}.}.
Therefore, $A_0 (w)$ ($A_3 (w)$) is an even (odd) function of $w$, and the regularity condition leads to $A'_0 (0) = 0$ ($A_3 (0) = 0$).
From the parity invariance, $\Psi_+$ ($\Psi_-$) is either an even (odd) function of $w$ or an odd (even) function. Since (\ref{peq1}) is invariant under $(\Psi_+,\Psi_-,q)\to (\Psi_-,\Psi_+,-q)$, the assignment of the parity is a convention up to the sign of $q$.
In this demonstration, we assign even parity to $\Psi_+$, and the regularity condition we employ is $\Psi'_+(0) = 0$.

We use the DBI action of the D8-branes as $S_{A}$: 
\begin{align}
&S_A = -C \int d^4x  dw  U^{1/4}(w) \notag \\
      &\times \sqrt{H^3(w) \{H^2(w) - (2 \pi l_s^2 / N_c)^2
      [ (A'_0 (w))^2-(A'_3 (w))^2]\}} \,,
\end{align}
where $C = (2 \pi)^{-7} (4 \pi^{2}/3) l_s^{-11/2} \lambda^{1/4} (2 M_{\text{KK}})^{-1/4} N_f N_c$, and
the $A_\nu (w)$ here is $N_c$ times $A_\nu (w)$ in the standard convention.
Here, $M_\text{KK}$ is the energy scale of the theory, $\lambda$ is  the 't~Hooft coupling, $N_f$ is the number of flavors, and $l_s$ is the string length which does not show up in the end of the calculation.
The definitions of $U(w)$ and $H (w)$ are given in \cite{HRYY}.

\begin{figure}
\begin{center}
\includegraphics[width=4.6cm, angle=-90]{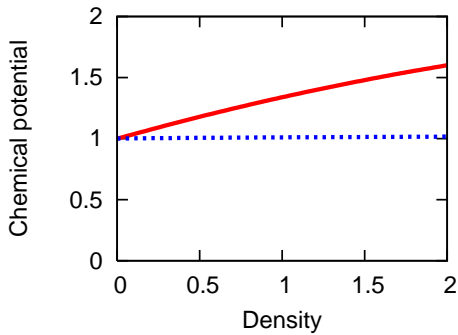}
\caption{
The EOS of the model (solid)
 compared with the one for the free baryons (dashed) given by $\mu = \sqrt{m_{n}^{2} + p_{F}^{2}}$, where $m_{n}$ and $p_{F}$ are the mass and the Fermi momentum of the baryons, respectively.
% $\mu/\mu_{n \to 0} = \sqrt{1 + (3 \pi^2 n/2)^{2/3}/\mu^2_{n \to 0}}$.
The horizontal axis is $n/n_0$, while the vertical axis is $\mu / m_{n}$.
}
\label{eos}
%\vspace{-8.8mm}
\end{center}
\end{figure}

We fix $M_\text{KK}$ and $\lambda$ in the same way as in \cite{SS}: $M_\text{KK} = 949$\,MeV and $\lambda N_c = 50$.
We consider $N_f = 2$ case, and use the normal nuclear matter density $n_0 = 0.16 \, \text{fm}^{-3}$
to scale the baryon number density $n$.
Figure~\ref{eos} shows the resultant EOS compared with the one for the free baryons.
The chemical potential $\mu$ increases as the density grows.
The increase of the chemical potential of free baryons is very tiny on the figure, whereas that of our dynamical baryons is considerably larger: this
can be understood as an effect of repulsive interactions among the baryons.

\begin{figure}
\begin{center}
\includegraphics[width=4.6cm, angle=-90]{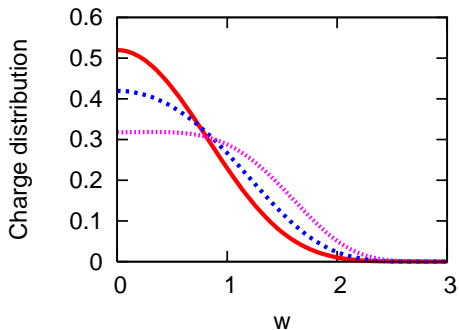}
\caption{
The baryon charge distribution $\rho (w)/n$ for $n/n_0 =$ $0.1$ (solid), $1$ (dashed) and $2$  (dotted). 
The boundary is located at $w = w_\text{max} \simeq 3.64/M_\text{KK}$.
The unit of $w$ on the figure is $1/M_\text{KK}$.
}
\label{dist}
%\vspace{-8.8mm}
\end{center}
\end{figure}
Figure~\ref{dist} shows the density dependence of the baryon charge distribution $\rho (w)$, which is indeed regular.
The distribution shifts to the boundary side as the density grows.
This implies that taking account of the distribution of the baryon charge along the radial direction becomes more important in the higher density region.

We expect that our approach is applicable to holographic descriptions of various systems of fermions~\cite{Hartnoll:2011fn}.

We thank 
D. K. Hong, Y. Kim, M. Rho, T. Sakai, H. Suganuma and P. Yi
for useful discussions and S. Hartnoll for the comment on the first version.
This work was initiated at the workshop Dense11
 at YITP.
This work is supported in part by the JSPS Grant-in-Aid (GIA) $\sharp$ 22224003 and $\sharp$ 24540266 (M.H.) and $\sharp$ 23654132 (S.N.), the GIA for Scientific Research on Innovative Areas No. 2104 (M.H. and S.N.), Nagoya Univ. GCOE Program QFPU (M.H. and S.T.)  from MEXT. 

Note added: After the first version appeared at the arXiv, we were informed on a recent related work~\cite{Sachdev:2011ze} where the Fermi distribution of the boundary theory is introduced as an input to obtain the EOS.

%\vspace{-6mm}

\end{document}